\documentclass[12pt]{article}
\usepackage{amssymb,latexsym}
\usepackage[mathscr]{eucal}

\newcommand{\g}{{\mathfrak g}}

\newcommand{\F}{{\mathcal F}}

\newcommand{\A}{{\mathcal A}}

\newcommand{\Z}{{\mathbb Z}}
\newcommand{\R}{{\mathbb R}}
\newcommand{\C}{{\mathbb C}}

\newcommand{\beq}{\begin{equation}}
\newcommand{\eeq}{\end{equation}}
\newcommand{\beqa}{\begin{eqnarray}}
\newcommand{\eeqa}{\end{eqnarray}} 
\newcommand{\tr}{\mbox{\rm Tr}\,} \newcommand{\uno}{\mbox{1
\kern-.59em {\rm l}}}

\newcommand{\nn}{\nonumber}
\newcommand{\be}{\begin{equation}}
\newcommand{\ee}{\end{equation}}
\newcommand{\bea}{\begin{eqnarray}}
\newcommand{\eea}{\end{eqnarray}}

\begin{document}
\begin{titlepage}
\begin{flushright}
{ROM2F/2004/05}\\
\end{flushright}
\begin{center}

{\large \sc Matone's Relation in the Presence of Gravitational Couplings}\\

\vspace{0.2cm}
{\sc Rainald Flume}\\
{\sl Physikalisches Institut der Universit\"at Bonn\\
Nu{\ss}allee 12, D--53115 Bonn, Germany} \\
{\sc Francesco Fucito}\\
{\sl Dipartimento di Fisica, Universit\'a di Roma ``Tor Vergata'',
I.N.F.N. Sezione di Roma II\\
Via della Ricerca Scientifica, 00133 Roma, Italy}\\
{\sc  Jose F. Morales}\\
{\sl Laboratori Nazionali di Frascati \\
P.O. Box, 00044 Frascati, Italy }
and\\
{\sc Rubik Poghossian}\\
{\sl Yerevan Physics Institute\\
Alikhanian Br. st. 2, 375036 Yerevan, Armenia}\\
\end{center}
\vskip 0.5cm
\begin{center}
{\large \bf Abstract}
\end{center}
{The prepotential in $N=2$ SUSY Yang-Mills theories enjoys
remarkable properties. One of the most interesting is its relation
to the coordinate on the quantum moduli space $u=\langle \tr
\varphi^2 \rangle$ that results into recursion equations for the
coefficients of the prepotential due to instantons. In this work we show, with an
explicit multi-instanton computation, that this relation holds
true at arbitrary winding numbers. Even more interestingly we show
that its validity extends to the case in which gravitational
corrections are taken into account if the correlators are suitably
modified. These results apply also to the cases in which matter in
the fundamental and in the adjoint is included. We also check that
the expressions we find satisfy the chiral ring relations
for the gauge case and
compute the first gravitational correction. }

\par    \vfill
\end{titlepage}
\addtolength{\baselineskip}{0.3\baselineskip}
\setcounter{section}{0}
\section{Introduction}
The chiral algebra of operators $O_m=\tr  \varphi^m$ in $N=2$
globally supersymmetric Yang-Mills (SYM from now on) gauge
theories in four space time dimensions has been the subject of intense research
in recent years. $\varphi(x)$ is the complex scalar field in the $N=2$
supersymmetric multiplet. One of the most remarkable results is the relation
between the expectation value $u=\langle \tr  \varphi^2 \rangle$\cite{Matone:1995rx}
and the $N=2$ prepotential
\beq
u(a)=i\pi(\F(a)-{1\over 2} a{\partial\over\partial a}\F(a))
\label{matrel} 
\eeq 
where $a$ is the v.e.v of the scalar field.
$u(a)=\sum_{k=0}^\infty {\cal G}_k (\Lambda/a)^{4k}a^2$, in fact,
obeys a non-linear differential equation that leads to a recursion
relation among the ${\cal G}_k's$. In turn, expanding
$\F(a)=\sum_{k=0}^\infty \F_k (\Lambda/a)^{4k}a^2$ and using
(\ref{matrel}) we obtain ${\cal G}_k=2\pi i k\F_k$ i.e. the
explicit expression of the prepotential.
We consider the $N=2$ theory with $SU(2)$ as underlying gauge group.
The generalization of (\ref{matrel}) for models with higher rank gauge 
group has also been considered \cite{Sonnenschein:1995hv}.
It was argued in \cite{Howe:1996pw} that (\ref{matrel}) is a consequence
of the Ward identities of (broken) superconformal invariance.

Inverting the functional dependence, i.e., taking $a=a(u)$ instead of $u=u(a)$,
one derives from (\ref{matrel}) for $a(u)$ and $a_D(u)=\partial \F/\partial a$
the second order differential equation
\beq
{\partial^2\over\partial u^2 }a_D(u)+V(u)a_D(u)=0
\label{matrel1}
\eeq
with $V(u)=-a^{-1}_D(u)\partial^2 a_D(u)/\partial u^2=a^{-1}(u)\partial^2 a(u)/\partial u^2$.
(\ref{matrel}) (or (\ref{matrel1})) alone does not allow for a determination of $u(a)$
and hence $\F(a)$. The necessary complementing information can be extracted from the Seiberg-Witten
curves \cite{sw}. This leads in the $SU(2)$ case to the determination $V(u)=-1/[4(1-u^2)]$.
Furthermore in \cite{Chan:1999gj} $u(a)$ was expressed in terms of theta functions. Both 
approaches, that of \cite{Matone:1995rx} and that of \cite{Chan:1999gj}, give rise to recursion relations
for the expansion coefficients of the prepotential  $\F(a)$.

A more recent achievement is the derivation of relations for expectation values of operators in 
the quantum chiral ring \cite{Cachazo:2002ry}. Those have been obtained by exploiting 
equations deduced from the Konishi anomaly.

In this paper we want to undertake the computation of the expectation values of the operators $O_m$
within the framework of microscopic instanton calculus. The basic references to rely on for this aim
are \cite{ft,DKM}. In the first of these references the first computation for $O_2$ was carried out.
In the second the general instanton configuration with its associated zero modes (and the classical
solution for the scalar field) has been constructed. We will also make use of the technique 
of equivariant localization \cite{berline} which, in the context of instanton calculus
has been first advanced in \cite{Bellisai:2000bc,Flume:2001kb} and has been 
put on a solid ground by \cite{Hollowood:2002ds} who observed that the deformation of 
the instanton configuration into the non commutative realm is a convenient device for the resolution
of the singularities of the corresponding moduli space. See also \cite{nak1,Nekrasov:1998ss}
for a more mathematical oriented description. The localization employed is akin to that of \cite{at}
who evaluated the equivariant Euler character. The method is discussed at length in \cite{Bruzzo:2002xf}
and extended to supermanifolds in \cite{Bruzzo:2003rw}. The final ingredient to be used is found in 
\cite{Nekrasov:2002qd} where 
it was proposed to use an extended vector field in the localization procedure which is related to the 
unbroken $U(1)$ symmetries and to space time rotations. The introduction of the latter gives rise
to a discrete set of critical points and makes therefore the task of evaluating the localized integrals
feasible. See also \cite{Flume:2002az,Bruzzo:2002xf} for related work.
It was also suggested in \cite{Nekrasov:2002qd} that the parameters of the space time rotations,
to be called below $\epsilon_1, \epsilon_2$, have a physical meaning in the sense that they are associated
with gravitational couplings.

We will also derive the above mentioned relations of the chiral operator ring and discuss some 
of the modifications due to gravitational couplings. We find in this way an independent
verification of (\ref{matrel}) which was partially checked in \cite{ft,Dorey:1996zj}.
More importantly we find a deformation of the scalar field, $\tilde\varphi$ which for 
$\epsilon_1, \epsilon_2\to 0$ falls back to $\varphi$ and preserves for $\epsilon_1, \epsilon_2\ne 0$
the chiral property. Moreover (\ref{matrel}) and (\ref{matrel1}) hold for $\tilde\varphi$ in
presence of gravitational couplings. It remains as an important open  problem 
to find the explicit expression for $V(u)$ for $\epsilon_1, \epsilon_2\ne 0$.

The survival of (\ref{matrel}) in presence of gravitational couplings may not be too surprising.
In fact in \cite{Howe:1996pw} it was argued that the l.h.s. in (\ref{matrel}) should emerge from a 
gauge invariant operator of canonical dimension two. With or without $\epsilon_1, \epsilon_2$ 
there is no other operator with this specification but $O_2$. 

The plan of the paper is the following: in section 2 we give some
preliminaries. In section 3 we discuss the classical solution for
the scalar field in the presence of a deformation
whose parameters $\epsilon_1, \epsilon_2$ will be introduced in due time.
In section 4 we compute correlators of the type $\langle \tr
\tilde{\varphi}^m \rangle$ and the chiral ring relation
\cite{Cachazo:2002ry} with its first gravitational correction.
The results of Section 3
and 4 are to be compared with those obtained in
\cite{Losev:2003py}. We stress that the object which satisfies the
Matone's relation in presence of the parameters $\epsilon_1,
\epsilon_2$ is $\tilde\varphi$ which is not a solution of the
Euler-Lagrange equations of motion.  
In appendix A we discuss the meaning of
computing vacuum expectation values in the non commutative case
and compare with the commutative case using the semiclassical
approximation.

\section{Preliminaries \label{intro}}
\setcounter{equation}{0}

\subsection{A brief reminder of the ADHM construction}

The starting point in the ADHM construction of $SU(n)$ self-dual
gauge connections with winding number $k$ is the $[2k+n]\times [2k]$
ADHM matrix
\beq \label{salute1} \Delta=\Delta_0-b z=\pmatrix{ w\cr
a^\prime-z }
\eeq
with
\beq w\equiv \pmatrix{J & I^\dagger}
\quad\quad a^\prime=\pmatrix{ B_1 & -B_2^\dagger\cr B_2 & B_1^\dagger}
\quad\quad z= \pmatrix{ z_1 & -\bar z_2\cr z_2 & \bar z_1}
\label{salute2}
\eeq
Here $J,I^\dagger$ and $B_i$ are $[n]\times [k]$ and $[k]\times
[k]$ matrices respectively.  These matrix elements can be taken to
be the coordinates ${\bf m}=\{B_1,B_2,I,J\}$ of a
$2k^2+2kn$ complex dimensional hyperk\"{a}hler manifold $\C^{2k^2+2kn}$.
$z_1, z_2$ are the complex coordinates
of the Euclidean space-time.
The form of $b=\pmatrix{0 \cr 1}$
has been fixed by exploiting the symmetries of the ADHM
construction. From now on, for the sake of simplicity, anytime we
write the matrix $z$ we intend it is multiplied by the $[k]\times
[k]$ unit matrix.

The self-dual gauge connection is
\beq A_\mu=\bar U(x)\partial_\mu U(x) \label{gaugeconn}
 \eeq
with $U$ a $[2k+n]\times [n]$ matrix in the
kernel of $\Delta$
\beq
\bar{\Delta}\,U=0=\bar{U}\,\Delta
\label{kerU}
\eeq
Self-duality of the field strength coming from (\ref{gaugeconn})
requires the matrix $\Delta$ to obey the constraint \beq
\bar{\Delta}\Delta = f^{-1}_{k\times k} {\bf 1}_{[2]\times[2]}
\label{bos} \eeq with $f_{k\times k}$ an invertible $[k]\times
[k]$ matrix. Substituting (\ref{salute1}) in (\ref{bos}) this
condition translates into the so called ADHM constraints
\beqa\label{mommap} f_{\C}&=&[B_1,B_2]+IJ=0\ \ ,
\nonumber\\
f_{\R}&=&[B_1,B_1^\dagger]+[B_2,B_2^\dagger]+II^\dagger-J^\dagger
J=0 \label{adhmc}\eeqa
 In the non commutative case in which we allow $[z_1,\bar
z_1]=-\zeta/2, [z_2,\bar z_2]=-\zeta/2$ \cite{Nekrasov:1998ss},
the above condition becomes $f_{\C}=0, f_{\R}=\zeta$.

The transformations
\beqa
\pmatrix{ w\cr a^\prime-z }\to \pmatrix{
T_{ a}\, w\, T_{ \phi}^{-1}T_{\epsilon_+}\cr T_{ \phi} T_{
\epsilon_-}\, (a^\prime-z)\, T_{ \phi}^{-1}T_{\epsilon_+} }
\label{uk}
\eea
with $T_\phi=e^{i\phi} \in U(k)$, $T_{ a} \in SU(n)$ and $ T_{
\epsilon_\pm}=e^{i\epsilon_\pm\sigma_3}$ leave the ADHM constraints
(\ref{adhmc}) invariant.
The transformations $T_\phi$ reflect the redundancy in the  ADHM description
and do not change the gauge connection (\ref{gaugeconn}).
$T_{ a}$ implements global gauge transformations which in the context of
$N=2$ SYM can be taken to be diagonal $T_a=\exp{\rm diag}(a_1,\ldots,a_n)$.
$T_{\epsilon_\pm}$
generate rotations in the complex $z_1, z_2$ planes with angles
$\epsilon_1=\epsilon_++\epsilon_-$ and $\epsilon_2=\epsilon_+-\epsilon_-$.

After imposing the $3k^2$ real constraints $f_\C=0, f_\R=\zeta$ we are left with a
manifold $M^\zeta$ of real dimension $k^2+4kn$.
We define the ADHM manifold ${\cal
M}_k^\zeta$ as the $U(k)$-quotient \beq {\cal M}_k^\zeta=M^\zeta/U(k)
\label{m}
\eeq
with $U(k)$ acting as in (\ref{uk}). For generic $\zeta$, ${\cal
M}_k^\zeta$ is a smooth manifold of dimension $4kn$ \cite{nak1}.

In the absence of v.e.v's, $\epsilon$ deformations and in the
commutative case, the solutions of the classical Euler-Lagrange
equation of motion of $N=2$ SYM for gauginos $\psi$ and scalar
field $\varphi$ can be written as
\beqa
\psi(x) &=& \bar{U}\left( {\cal M} f \bar{b}-b f \bar{{\cal M}}\right)U \nonumber\\
i \varphi_{\rm ferm}(x) & =&
\bar{U}\,{\cal{M}} f\bar{\cal M}U
+  \bar{U}  \pmatrix{ 0_{[n]\times [n]} & 0_{[n]\times[2k]}
\cr 0_{[2k]\times[n]} &\A^\prime_{f[k]\times[k]}\otimes
1_{[2]\times[2]}}  U.
\label{An}
\eeqa
The $[k]\times [k]$
matrix $\A_f^\prime=\A^\prime_f$ obeys ${\bf
L}\A_f^\prime=\Lambda_f$ where \beq \Lambda_f={1\over 2}
\left(\bar{\cal M}{\cal M} - (\bar{\cal M}{\cal M})^T\right)
\eeq
and the operator ${\bf L}$ is defined as
\beq {\bf L}\cdot
\Omega = \{II^\dagger+J^\dagger J,\Omega\} +
\sum_{m=1,2}[B_m,[B^\dagger_m,\Omega]]+[B^\dagger_m,[B_m,\Omega]].
\label{L} \eeq Finally ${\cal M}=\pmatrix{\mu \cr {\cal
M}^\prime}$ is a constant $[2k+n]\times [2k]$ matrix of
Grassmanian collective coordinates,the fermionic analogue of the
matrix $\Delta_0$ introduced in (\ref{salute1}). It satisfies the
fermionic ADHM constraint
\beq \bar{\Delta}_0{\cal M} = \bar{\cal
M}\Delta_0 \label{amma} \eeq Given two arbitrary variations of the
ADHM data (\ref{salute2}) we define their scalar product as \beq
\langle \delta_1 \Delta_0,\delta_2\Delta_0\rangle= \tr_k\Re
(\delta_1\bar \Delta_0\cdot\delta_2\Delta_0).
\label{scaprod}
\eeq
Let us point out that this scalar product (\ref{scaprod}) in the
moduli space is induced by the conventional definition of the
scalar product of gauge zero modes in Euclidean space time.
Finally we introduce a
$U(k)$-covariant derivative $D=d+C$ with $C$ determined by the
condition that covariant derivatives in the moduli space are
orthogonal to all infinitesimal $T_\phi\in U(k)$ variations \beq
0=\langle D\Delta_0,\delta_\phi\Delta_0 \rangle =\tr_k[\phi ({\bf
L}\,C-{\bf X})] \label{cd} \eeq with \beq {\bf X}=\tr\,
[B_1^\dagger,dB_1]+[B_2^\dagger,dB_2] -dI I^\dagger+J^\dagger dJ\
-{\rm h.c.} . \eeq
It follows from (\ref{cd}) that ${\bf
L}\,C={\bf X}$. The connection $C$ is the analog of the gauge transformation
needed to impose the background gauge condition to gauge zero
modes.
If we put the covariant derivative $D\Delta_0$ in
place of  ${\cal M}$ (\ref{amma}), the fermionic constraint is
automatically satisfied. This allows to identify fermionic zero
modes ${\cal M}$ with the one-form $D\Delta_0$
\cite{Bellisai:2000bc,Flume:2001kb}.
 This can be used to rewrite the fermion bilinear in (\ref{An}) as
 \be
 \bar{U}\,{\cal{M}} f\bar{\cal M}U=(D\bar{U})(DU)
\label{ummu}
 \ee
with
\beq
D\bar U=-\bar U(D\Delta_0)f\bar\Delta,\qquad D U=-\Delta f (D\Delta_0)^\dagger U.
\label{closedform2}
\eeq
following\footnote{In (\ref{closedform2}) we have neglected the gauge transformation terms since we
think of always dealing with gauge invariant forms.}  
from (\ref{kerU}) \cite{DKM}. This observation will be helpful in what follows.

\subsection{Equivariant Forms and the Localization Formula}

 In this subsection we briefly review the localization
formula. See \cite{Bruzzo:2002xf,Bruzzo:2003rw} for a more
detailed discussion in this context. We also discuss the
geometrical setting in which the localization formalism will be
applied. We have seen that $M^\zeta$ is acted upon by a Lie group
$G\equiv U(1)^n\times U(1)^2$ with Lie algebra $\g$.  For every
$\xi\in\g$  we denote by $\xi^*=\xi^s
\,T_{s}^i\frac{\partial}{\partial m^i}$ the \emph{fundamental
vector field} associated with $\xi$, where the
$\xi^s$ are the components of $\xi$
in some chosen basis of $\g$, and the $T_s^i$
are the generators of the action with
$\xi^s T_s^i=\delta_{\xi^s}\, m^i$.
$\xi^*$ is the vector field that
generates the one-parameter group $e^{t \xi}$ of transformations
of $M^\zeta$ \beq \xi^* \pmatrix{w\cr a^\prime}\equiv \delta_\xi
\pmatrix{w\cr a^\prime}\equiv i_\xi \pmatrix{dw\cr da^\prime}=
\pmatrix{aw+w \epsilon_+ \sigma_3\cr \epsilon_- \sigma_3
a^\prime+\epsilon_+ a^\prime \sigma_3} \label{xis} \eeq given by
(\ref{uk}).
We have seen in (\ref{m}) that the moduli space ${\cal M}_k^\zeta$
is the space of the $U(k)$ orbits on $M^\zeta$. In order to make
the variations (\ref{xis}) orthogonal to these orbits we have to
use the previously introduced $U(k)$ connection $C$. These
orthogonal variations are given by
\beq
\tilde\xi^* \pmatrix{w\cr
a^\prime}\equiv \delta_{\tilde\xi} \pmatrix{w\cr a^\prime}\equiv
i_\xi \pmatrix{Dw\cr Da^\prime}=\pmatrix{aw+w (-\phi+ \epsilon_+
\sigma_3)\cr [\phi,a^\prime]+\epsilon_- \sigma_3 a^\prime+\epsilon_+
a^\prime \sigma_3}
\label{xis1}
\eeq
with $\phi\equiv i_{\xi}
C$. In order to apply the localization theorem we introduce the
equivariant differential $d_{\xi}$ by letting

\beq d_\xi \equiv d+i_\xi
\label{der}
\eeq
Acting twice on an
equivariant form $\alpha$ (a form satisfying $g \alpha=\alpha$ for
all $g\in \g$) one finds \beq d_\xi^2 \alpha=(d i_\xi+i_\xi
d)\alpha={\cal L}_\xi\alpha=0 \ \eeq where ${\cal L}_\xi$ is the
Lie derivative. Thus the space of equivariant forms becomes a
differential complex. A form is said to be equivariantly closed if
it is equivariant and satisfies $d_\xi\alpha=0$. The condition
that a form is equivariantly closed implies that under certain
rather general conditions its top form is exact outside the zeros
of $\xi^*$ \cite{berline}, suggesting that the integral localizes
around the critical points.

It is now time to discuss the action of $N=2$ SYM on the moduli space.
In presence of a
v.e.v. for the scalar field, in addition to the piece $8\pi^2
k/g^2$ ($g$ is the gauge coupling constant and $k$ the winding
number) the action becomes dependent upon the moduli.
It can be evaluated by computing the norm of the zero mode
${\mathcal D}\varphi$, where ${\mathcal D}$ is the covariant
derivative with respect to
the gauge connection \cite{DKM,Dorey:2002ik}. An alternative representation
of $S$ is given as a BRST variation, $Q$, of a one-form,
$\Omega=\langle \xi^*,D m \rangle$.
After identifying $Q$ with $D+i_{\xi}$
we get $S=d_\xi\Omega$
\cite{Bellisai:2000bc,Flume:2001kb,Flume:2002az} that implies in
particular that $d_\xi S={\cal L}_\xi \Omega=0$, i.e. $e^{-S}$ is
an equivariantly closed form. Moreover at the fixed point $S=0$.

The equations for the fixed points of the vector field action
express the condition that a point of $M^\zeta$ acted upon by
$\xi^*$ is left invariant up to a $U(k)$ transformation \beqa
\pmatrix{ T_{ a}\, wT_{\epsilon_+}\cr T_{\epsilon_-}\, a^\prime
T_{\epsilon_+}} =\pmatrix{ w\, T_{ \phi}\cr T^{-1}_{ \phi}
a^\prime T_{ \phi}}. \label{uk1} \eea This is the same as setting
to zero the components of $\tilde\xi^*$ in (\ref{xis1}). Moving
along the $U(k)$ orbit (\ref{uk1}) remains valid for different
$T_{ \phi}$. For simplicity we choose to solve (\ref{uk1}) at that
point of the orbit where $T_{ \phi}={\rm
diag}(e^{i\phi_1},\ldots,e^{i\phi_k})$
\bea
&&a J+J(-\phi+\epsilon_+)=(\phi_{I}-a_\alpha-\epsilon_{+})J_{\alpha I}=0\nn\\
&&I a-(\phi+\epsilon_{+})I=(\phi_I-a_\alpha+\epsilon_{+})I_{I \alpha}= 0\nn\\
&&[\phi,B_\ell]+\epsilon_{\ell}
B_\ell=(\phi_{IJ}+\epsilon_{\ell})B_{\ell J I}=0 \label{fixe} \eea
with $\ell=1,2$ and $\phi_{IJ}=\phi_{I}-\phi_J$. Integrals over
the ADHM manifold will localize around the $\xi^*$-fixed points
i.e. the solutions of (\ref{fixe}). The critical points are in one
to one correspondence to the partitions of $n$ integers
$k_{\alpha}$ with $\alpha=1,\ldots,n$ with $\sum_\alpha
k_\alpha=k$ the total winding number \cite{Nekrasov:2002qd}
\footnote{It is here assumed that the parameter measuring the non
commutativity of the instanton moduli manifold is non zero.}. In
the picture of the ADHM construction as a system of $k$ D(-1)
branes superposed on $n$ D3 branes, this corresponds to
distributing in all possible ways the integer $k$ $D(-1)$-branes
between the $n$ D3 branes and then consider all the possible
partitions of the resulting $k_\alpha$ 's. $\phi_{I_\alpha}\equiv
\phi_{\alpha;( ij)}$ is the $U(k)$-parameter associated to the box
in the i-th row and j-th column in the $\alpha$-th Young diagram
\beq \phi_{I_\alpha}\equiv \phi_{\alpha;(i,j)}=
a_\alpha-\epsilon_+ +(i-1)\epsilon_1+(j-1)\epsilon_2
\label{ficritical} \eeq with $i,j\in Y_\alpha$. Solutions to
(\ref{fixe}) are found by setting to zero all components in
$B_\ell,I,J$ except for elements of the form
$(B_1)_{(i,j)(i-1,j)}, (B_2)_{(i,j)(i,j-1)}$ and $I_{1,1},
I_{k_1+1,2},\ldots,I_{k_{n-1}+1,n}$. These elements are later
completely fixed by imposing ADHM constraints. This implies in
particular that the vector field $\tilde\xi^*$ has only isolated
zeroes. Assuming that $\alpha$ is equivariantly closed and that
$\xi\in\g$ is such that the vector field $\xi^*$ has only isolated
zeroes, $x_0$, we can state the localization theorem
\beq \int_M
\alpha=(-2\pi)^{n/2}\sum_{x_0}{\alpha_0(x_0)\over {\rm
det}^{1\over 2}\, {\cal L}_{x_0}}
\label{locth}
\eeq
where the
map ${\cal L}_{x_0}:T_{x_0} M\to T_{x_0} M$  is defined as
\beq\label{unoapp}
{\cal L}_{x_0}(v) =[\xi^*,v]= -\xi^\alpha\,
v^i\left(\frac{\partial T_\alpha^j}{\partial
m^i}\right)_{x_0}\frac{\partial}{\partial m^j},
\eeq
(which makes
sense because at the critical points the components of the
fundamental vector field vanish, $\xi^\alpha T_\alpha^i(x_0)=0$).
As first appreciated in \cite{Nekrasov:2002qd} this localization
takes place in the evaluation of the centered partition function
of $N=2$ SYM. Taking $\alpha=e^{-S}$, with $S$ the multi-instanton
action, the application of (\ref{locth}) at winding number $k$
leads to \cite{Flume:2002az,Bruzzo:2002xf} \beqa {\cal
Z}_k=\sum_{x_0} {1\over {\rm det}
 \hat{{\cal L}}_{x_0}} &=&\sum_{\{Y_\alpha; \sum_\alpha\vert
Y_\alpha\vert=k \}}
\prod_{\alpha,\beta=1}^n \prod_{s\in Y_{\alpha}}{1\over
E_{\alpha\beta}(s)(2\epsilon_+-E_{\alpha\beta}(s))}
\label{generalsdet}
\eeqa with
\beqa E_{\alpha\beta}(s) &=&
a_{\alpha\beta}-\epsilon_1 h_\beta(s)+\epsilon_2(v_\alpha(s)+1)
\eeqa
where $h_\beta(s)$ ($v_\alpha(s)$) denotes the
horizontal(vertical) distance from the box "s" till the upper (left)
end of the $\beta$($\alpha$)-diagram i.e.  the number of black
(white) circles in Fig.1.
From ${\cal Z}_k$ we can build the generating function ${\cal Z}=\sum_k{\cal Z}_kq^k$,
where $q=e^{2\pi i\tau}$ and $\tau$ is the coupling of $N=2$ SYM.

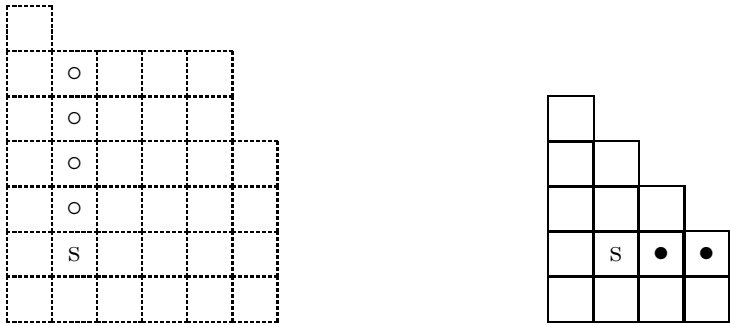
\begin{figure}[ht]
\label{figura} \setlength{\unitlength}{2mm}
\begin{center}
\begin{picture}(60,25)(-5,-20)
\put(0,0){\dashbox{.2}(3,3)}
\put(0,-3){\dashbox{.2}(3,3)}\put(3,-3){\dashbox{.2}(3,3){$\circ$}}\put(6,-3){\dashbox{.2}(3,3)}
\put(9,-3){\dashbox{.2}(3,3)}\put(12,-3){\dashbox{.2}(3,3)}
\put(0,-6){\dashbox{.2}(3,3)}\put(3,-6){\dashbox{.2}(3,3){$\circ$}}\put(6,-6){\dashbox{.2}(3,3)}
\put(9,-6){\dashbox{.2}(3,3)}\put(12,-6){\dashbox{.2}(3,3)}
\put(0,-9){\dashbox{.2}(3,3)}\put(3,-9){\dashbox{.2}(3,3){$\circ$}}\put(6,-9){\dashbox{.2}(3,3)}
\put(9,-9){\dashbox{.2}(3,3)}\put(12,-9){\dashbox{.2}(3,3)}\put(15,-9){\dashbox{.2}(3,3)}
\put(0,-12){\dashbox{.2}(3,3)}\put(3,-12){\dashbox{.2}(3,3){$\circ$}}\put(6,-12){\dashbox{.2}(3,3)}
\put(9,-12){\dashbox{.2}(3,3)}\put(12,-12){\dashbox{.2}(3,3)}\put(15,-12){\dashbox{.2}(3,3)}
\put(0,-15){\dashbox{.2}(3,3)}\put(3,-15){\dashbox{.2}(3,3){s}}\put(6,-15){\dashbox{.2}(3,3){}}
\put(9,-15){\dashbox{.2}(3,3){}}\put(12,-15){\dashbox{.2}(3,3){}}
\put(15,-15){\dashbox{.2}(3,3){}}
\put(0,-18){\dashbox{.2}(3,3)}\put(3,-18){\dashbox{.2}(3,3)}
\put(6,-18){\dashbox{.2}(3,3)} \put(9,-18){\dashbox{.2}(3,3)}
\put(12,-18){\dashbox{.2}(3,3)}\put(15,-18){\dashbox{.2}(3,3)}
\put(36,-6){\framebox
(3,3)}
\put(36,-9){\framebox (3,3)}\put(39,-9){\framebox
(3,3){}}
\put(36,-12){\framebox (3,3)}\put(39,-12){\framebox
(3,3){}}\put(42,-12){\framebox (3,3)}
\put(36,-15){\framebox (3,3)}\put(39,-15){\framebox
(3,3){s}}\put(42,-15){\framebox (3,3){$\bullet$}}
\put(45,-15){\framebox
(3,3){$\bullet$}}
\put(36,-18){\framebox (3,3)}\put(39,-18){\framebox
(3,3)}\put(42,-18){\framebox (3,3)} \put(45,-18){\framebox
(3,3)}
\end{picture}
\caption{Two generic Young diagrams denoted by $Y_\alpha$ (dotted line) and
$Y_\beta$ (solid line) in the main text. }
\end{center}
\end{figure}

\section{The scalar field in presence of a deformation}
\setcounter{equation}{0}

In this section we will study how the classical solution for the
scalar field in (\ref{An}) gets modified in presence of the v.e.v.'s,
$a_\alpha$, and the rotations given from $\epsilon_{\pm}$.
The solution thus found is the general solution to the deformed equations of motion. 
In the next section we will
see how its zero form part can be evaluated at the critical points (\ref{fixe}).

It is well known that the covariant derivative of the part of the
scalar field obeying to the homogeneous Euler-Lagrange equation
\footnote{And to the boundary conditions $\lim_{x\to\infty}\varphi=a$.}
(turning off the fermionic sources), $Z^a_\mu$, is a zero mode satisfying
\beq \nabla^\mu Z^a_\mu=0,\quad \nabla_{[\mu}
Z^a_{\nu]}=(\nabla_{[\mu} Z^a_{\nu]})^{dual} \label{zeromod} \eeq
with 
\beq Z^a_\mu\equiv 2\, Im \{
\bar{U}\tilde\xi_a^*\Delta_0 \bar{\sigma}_\mu f \bar{b}U\}
\label{zeromod1} 
\eeq 
where $\tilde\xi_a^*$ is obtained from
(\ref{xis1}) setting $\epsilon_\pm=0$. To find the solution,
$Z_\mu$, in presence also of the rotations $\epsilon_\pm$, we take
as an ansatz (\ref{zeromod1}) with $\tilde\xi_a^*$ replaced by $\tilde\xi^*$
given in (\ref{xis1}).

With this in mind, following Appendix (C.1) of \cite{DKM}, and
writing
\beqa
\xi^*\Delta_0 &=&\Delta(-\phi+\epsilon_+
\sigma_3)+\pmatrix{a&0\cr 0& \phi+\epsilon_- \sigma_3}\Delta-
\pmatrix{0\cr \epsilon_+ z\sigma_3+\epsilon_- \sigma_3 z} \nn
\eeqa
one finds
\beqa Z_\mu &=& 2\, Im\{ \bar{U}
 \left[   \pmatrix{a&0\cr 0&
\phi+\epsilon_- \sigma_3}\Delta- \pmatrix{0\cr \epsilon_+
z\sigma_3+\epsilon_- \sigma_3 z} \right]\bar{\sigma}_\mu f \bar{b}
U \} \nn\\
&=& 2\, Im\{ \bar{U}
   {\cal A}_{\rm bos} \Delta \bar{\sigma}_\mu f \bar{b}
U \}-\Omega_\lambda^\nu x^\lambda F_{\mu\nu} =D_\mu \varphi_{\rm
bos}-\Omega_\lambda^\nu x^\lambda F_{\mu\nu}
\label{zmu}
\eeqa
 with $\varphi_{\rm bos}=\bar{U}\,{\cal A}_{\rm bos}\,U $ and
\beq {\cal A}_{\rm bos}=\pmatrix{a&0\cr 0& \phi+\epsilon_-
\sigma_3},\qquad \Omega^\mu_\nu=\pmatrix{0&-\epsilon_1&0&0\cr
\epsilon_1&0&0&0\cr0&0&0&-\epsilon_2\cr 0&0&\epsilon_2&0}
\label{a}
\eeq
in agreement with \cite{Losev:2003py}. The complete
solution for the scalar field in presence of v.e.v's and
gravitational fields is obtained summing $\varphi_{\rm bos}$ to
(\ref{An}), i.e. $\varphi=\varphi_{\rm ferm}+\varphi_{\rm bos}$.

Notice that the $U(k)$ field $\phi\equiv i_\xi C$ satisfies
the equation
\beqa
 {\bf L}\,\phi &=&{\bf
L}\,i_\xi\, {\bf L}^{-1} {\bf X}= i_\xi {\bf X}\equiv
\Lambda_{B} \nn\\
&\equiv & \sum_{\ell=1,2}i\epsilon_\ell[B_\ell,B^\dagger_\ell]+
iJ^\dagger(a+\epsilon_+) J+ i I(a+\epsilon_+)I^\dagger
\label{lambdaC}
\eeqa
and therefore it matches the standard
solution at $\epsilon_{\pm}=0$.

\section{Correlators containing the scalar fields}
\setcounter{equation}{0}
\subsection{The zero form part of the scalar field}
We are now ready to discuss the application of (\ref{locth}) to
the computation of correlators containing scalar fields. The
correlator we are interested in is $\langle{\cal
O}(x)\rangle$ which is a composite of scalar fields.
In order to apply the localization
formula (\ref{locth}) we need to deal with equivariantly closed
form.
The right objects are $\langle \tr \tilde{\varphi}^m \rangle$
with $\tilde{\varphi}=\tilde{\varphi}_{\rm bos}+\varphi_{\rm
ferm}$ a closed equivariant form and
\beq \tilde{\varphi}_{\rm bos}= \bar{U} \delta_\xi U.
\eeq
 That $\tilde{\varphi}$ is equivariantly closed, i.e.
 $Q\tilde{\varphi}=(D+i_\xi)\tilde{\varphi}=0$ can be seen as follows
 \footnote{In the previous sections of the paper the forms that enter
(\ref{locth}) are all gauge invariant. In this case $Q$ coincides with (\ref{der}).}.
 First we recall  that in the absence of v.e.v. and gravitational
 backgrounds the scalar field (\ref{An}) is BRST closed i.e. $D\varphi_{\rm ferm}=0$.
 This implies that the only contributions to $Q\varphi_{\rm ferm}=i_\xi \varphi_{\rm ferm}$
 comes from fermion bilinears in (\ref{An}) since the contraction $i_\xi$ is trivial
 on zero forms. Moreover $i_\xi \A_f^\prime=0$ as follows
 from the fermionic ADHM constraint \label{amma}. Finally writing the fermion bilinears
 as in (\ref{ummu}) one finds
\beqa
Q\tilde{\varphi}&=&(D+i_\xi)(\varphi_{\rm ferm}+\tilde{\varphi}_{\rm bos})
= i_\xi (D\bar{U} DU)+D\bar{U}\delta_\xi U+\bar{U} D\delta_\xi U\nn\\
&=&\delta_\xi \bar{U} DU-D\bar{U} \delta_\xi U+D\bar{U}\delta_\xi U+\bar{U} D\delta_\xi U=0
\eeqa
where in the last equation we use $\delta_\xi \bar{U} U=-\bar{U} \delta_\xi U$.
We conclude that it is $\tilde{\varphi}$ rather than ${\varphi}$ the equivariantly
closed form suitable for localization.

The transformation rules for $U$ can be read from those of
$\Delta$ in (\ref{uk})
\beqa
\Delta^\prime (z_\epsilon)&=&
\pmatrix{T_a&0 \cr 0& T_\phi T_{\epsilon_-} } \Delta(z)  \,
\pmatrix{T_\phi^{-1}T_{\epsilon_+}& 0\cr 0&
T_\phi^{-1}T_{\epsilon_+}}
\nonumber\\
U^\prime(z)&=& \pmatrix{T_a&0\cr 0& T_\phi T_{\epsilon_-}}
U(z_{-\epsilon})
\label{transfinal}
\eeqa
Taking the infinitesimal
variation
\beqa \tilde{\varphi}_{\rm bos}=\bar{U}\delta_xi U(z)&=&
\bar{U}\pmatrix{ia_\lambda &0 \cr 0& i\phi+i\epsilon_- \sigma_3
}U(z)
-\bar{U}iz_\ell\epsilon_\ell{\partial\over\partial z_\ell}U(z)\nonumber\\
&=&\bar{U}(\A_{\rm bos}-iz_\ell\epsilon_\ell{\partial\over\partial
z_\ell})U(z)=\varphi_{\rm
bos}-i\bar{U}z_\ell\epsilon_\ell{\partial\over\partial
z_\ell}U(z).
\label{fi}
\eeqa
It will be shown in the following
subsection that it is $\tr \tilde\varphi^2=\tr(\tilde\varphi_{\rm
bos}+\varphi_{\rm ferm})^2$  rather than $\tr
\varphi^2=(\varphi_{\rm bos}+\varphi_{\rm ferm})^2$ alone the
quantity that obeys Matone's relation. Clearly the two quantities
coincide in the $\epsilon_{\pm}\to 0$ limit and therefore
the correlators $\langle \tr \varphi^m \rangle|_{\epsilon=0}$ can be
studied using theirs equivariantly closed deformation $\langle \tr
\tilde\varphi^m \rangle$.

According to the localization theorem $\langle \tr \tilde\varphi^m
\rangle$ reduces to its 0-form part evaluated at the critical
points times the inverse determinant appearing in
(\ref{generalsdet} ). The inverse determinant (\ref{generalsdet})
was already computed for $N=2, 2^*, 4$
\cite{Nekrasov:2002qd,Flume:2002az,Bruzzo:2002xf}. In our case,
since the action is zero at the critical points, $\alpha_0(x_0)$
is just the sum of the $m$-$th$ powers of the eigenvalues in
(\ref{fi}) i.e. $\alpha_0(x_0)=\tr\tilde\varphi_{\rm bos}^m$. We
remark that at a critical point, the matrix $\phi$ in (\ref{fi})
takes the values (\ref{ficritical}). In fact, substituting
(\ref{fixe}) in the l.h.s. of (\ref{lambdaC}) and using (\ref{L})
we obtain an identity.

To end the computation we now have to evaluate $U(z)$. To do so we
have to cope with a last problem: to use the localization formula
(\ref{locth}) the moduli space needs to be compactified and
desingularized \cite{nak1}. This deforms the real ADHM constraint
(\ref{mommap}) into $f_{\R}=\zeta$ as we said earlier. The
variables $z_1, z_2$ become operators in a Fock space ${\cal H}$.
The non commutative ADHM construction has been previously
studied in \cite{Nekrasov:1998ss,Furuuchi:1999kv,Chu:2001cx}.
In the following we will use the formalism elaborated in \cite{Furuuchi:1999kv}.
We first remind the reader that $U(z)$ is the
kernel of the operator matrix $\Delta^\dagger$. Then we rewrite $U(z)$ as
\beq \vert U(z)>=\pmatrix{\vert w>\cr \vert u>\cr \vert v>},
\qquad \matrix{\vert w>\equiv w(z_1,z_2)\vert 0,0>\cr \vert
u>\equiv u(z_1,z_2)\vert 0,0>\cr \vert v>\equiv v(z_1,z_2)\vert
0,0>} \label{states} \eeq $\vert u>, \vert v>\in {\cal H}^{\oplus
k}$, are vectors in $\C^k$ and in the Hilbert space ${\cal H}$,
$\vert w>\in {\cal H}^{\oplus n}$ is a vector in $\C^n$ and ${\cal
H}$. Now it is possible to show \cite{nak2}(we will check it
later) that the space spanned by (\ref{states}) is isomorphic to
the ideal \beq {\cal I}=\{ w(z_1,z_2) \vert w(B_1,B_2)=0\}
\label{ideal} \eeq The ideal ${\cal I}$ is given by all elements
of the form $w(z_1,z_2)=z_1^{k-1}z_2^{l-1}$ with $k,l$ running
over boxes outside the Young tableaux $Y_\alpha$. It is easy to
check that $B$-matrices associated to a tableaux $Y_\alpha$ belong
to the ideal (\ref{ideal}), i.e. $B_1^{k-1}B_2^{l-1}=0$ for all
$k,l\neq Y_\alpha$\footnote{This can also be easily checked from
the explicit solutions for $B_1, B_2$ given in
\cite{Flume:2002az}.}. If $\C^2[z_1,z_2]=\{z_1^{m-1}z_2^{n-1}\vert
m,n\in \Z_+\}$ is the ring of polynomials of $z_1, z_2$ and ${\cal
I}=\{z_1^{k-1}z_2^{l-1}\vert k,l\neq Y_\lambda\}$ then
$Y_\alpha=\C^2(z_1,z_2)/{\cal I}$ with ${\rm dim}Y_\alpha =k$. For
the sake of simplicity we now focus on the $U(1)$ case since the
$U(n)$ calculation follows straightforwardly from it.
\begin{figure}[ht]
\label{figura4}
\setlength{\unitlength}{2mm}
\begin{center}
\begin{picture}(20,20)(-5,-15)
\put(0,0)  {\framebox  (4,3){$\scriptstyle{z_2^4}$}}
\put(0,-3){\framebox
(4,3){$\scriptstyle{z_2^3}$}}\put(4,-3)
{\framebox(4,3){$\scriptstyle{z_1z_2^3}$}}
\put(8,-3){\framebox(4,3){$\scriptstyle{z^2_1 z_2^3}$}}
\put(0,-6){\framebox(4,3){$\scriptstyle{z_2^2}$}}\put(4,-6)
{\framebox(4,3){$\scriptstyle{z_1z_2^2}$}} \put(8,-6)
{\framebox(4,3){$\scriptstyle{z_1^2z_2^2}$}}

\put(0,-9){\framebox(4,3){$\scriptstyle{z_2}$}}\put(4,-9)
{\framebox(4,3){$\scriptstyle{z_1z_2}$}} \put(8,-9)
{\framebox(4,3){$\scriptstyle{z_1^2z_2}$}}
\put(12,-9){\framebox(4,3){$\scriptstyle{z_1^3z_2}$}}

\put(0,-12){\framebox(4,3){$\scriptstyle{1}$}}\put(4,-12)
{\framebox(4,3){$\scriptstyle{z_1}$}}\put(8,-12)
{\framebox(4,3){$\scriptstyle{z_1^2}$}}
\put(12,-12){\framebox(4,3){$\scriptstyle{z_1^3}$}}

\end{picture}
\caption{Notation for a generic Young diagram.}
\end{center}
\end{figure}

In Fig.2 we clarify our notation. The elements of ${\cal I}$ are
all the monomials that do not appear in the figure and are raised
to powers greater than those in the boxes.

We  remind the reader that a pair of indices $(i,j)$ is also
assigned to each box so that the non-trivial matrix elements in
the $B$-matrices are $(B_1)_{(i,j)(i-1,j)}, (B_2)_{(i,j)(i,j-1)}$.
$J$ is always zero while $I$ is a $k$-dimensional vector whose
only element different from zero is $I_1=\sqrt{k}$.\footnote{The
U(n) case is not much different: $J$ is still zero while $I$ is a
$[k]\times [n]$ matrix whose only elements different from zero are
$I_{1,1}, I_{k_1+1,2},\ldots,I_{k_{n-1}+1,n}$. The matrices $B_1,
B_2$ are the same as in the $U(1)$ case.} Let us now check the
isomorphism between the ideal ${\cal I}$ and the space spanned by
the states (\ref{states}). From $\Delta^\dagger U=0$, we find the
conditions \beqa
&&wI-(B_2-z_2)u+(B_1-z_1)v=0,\nonumber\\
&&(B_1^\dagger-\bar z_1)u+(B_2^\dagger-\bar z_2)v=0.
\label{sistema} \eeqa At the critical points, using the explicit
expression of $w$ given by (\ref{ideal}) one is lead to the
following form for the (1,1) and generic $(i,j)$ component in the
first matrix equation in (\ref{sistema}) \beqa
&&\sqrt{k} z_1^{k-1}z_2^{l-1}+z_2u_{(1,1)}-z_1v_{(1,1)}=0,\nonumber\\
&&(B_2)_{(i,j)(i,j-1)}u_{(i,j-1)}-z_1v_{(i,j)}+
(B_1)_{(i,j)(i-1,j)}v_{(i-1,j)}+z_2u_{(i,j)}=0.
\label{sistemacomponent} \eeqa (\ref{sistemacomponent}) is
satisfied by $u_{(i,j)}\sim z_1^{k-j}z_2^{l-i-1}, v_{(i,j)}\sim
z_1^{k-j-1}z_2^{l-i}$. The indices $i, j\in Y$ ($k, l\notin Y$)
denote those boxes which are inside (outside) a given Young
tableaux. The proportionality coefficients of these relations are
not important for our subsequent reasonings \footnote{The reader
can convince himself that these coefficients are zero whenever the
exponents of the monomials are negative. To determine these
coefficients also the second equation in (\ref{sistema}) is
obviously needed. It is a differential equation for $u, v$ once we
put $\bar z_{1,2}=\zeta/2
\partial/\partial z_{1,2}$.}.

Substituting the form of $U(z)$ thus obtained in (\ref{fi}) we
find that the operator $\delta_\xi$ acting on $U(z)$ has
eigenvalues
$\lambda_i=-ia_\alpha+i(k-1)\epsilon_1+i(l-1)\epsilon_2$.
 The Chern character is then defined by the sum
$\sum_i e^{i \lambda_i}$ over all eigenvalues $\lambda_i$. To this
end we introduce the notation $T_1=e^{i\epsilon_1},
T_2=e^{i\epsilon_2}, T_{a_\alpha}=e^{ia_\alpha}$. Then \beqa
\chi&=&\sum_{\alpha=1}^n\sum_{(k,l)\notin Y_\alpha}
T_{a_\alpha}T_1^{i-1}T_2^{j-1}=\sum_{\alpha=1}^n\bigg(\sum_{(m,n)\in\Z^2_+}T_{a_\alpha}
T_1^{m-1}T_2^{n-1} -\sum_{(i,j)\in Y_\alpha}
T_{a_\alpha}T_1^{i-1}T_2^{j-1}\bigg)\nn\\
&=& {1\over {\cal V}}\bigg[\sum_{\alpha=1}^n
T_{a_\alpha}+\sum_{\alpha=1}^n\sum_{(i,j)\in
Y_\alpha}T_{a_\alpha}T_1^{i-1}T_2^{j-1}(T_1-1)(T_2-1) \bigg]\label{lmnfinal} \\
&=&{1\over {\cal V}}\sum_{\alpha =1}^n\bigg\lbrace e^{i a_\alpha}+
\sum_{j_\alpha=1}^{k_{1_\alpha}}
\bigg[e^{i[a_\alpha+\epsilon_1(j-1)+\epsilon_2k^\prime_{j_\alpha}]}
-e^{i[a_\alpha-\epsilon_2(j-1)]}
-e^{i[a_\alpha+\epsilon_2j+\epsilon_1k^\prime_{j_\alpha}]}+e^{i[a_\alpha+\epsilon_2
j]}\bigg]\bigg\rbrace \nn \eeqa where $k_{j_\alpha}$
($k^\prime_{j_\alpha}$) are the number of boxes in the $j$-th row
(column) of the tableaux $Y_\alpha$. The sum over the index
$j_\alpha$ runs over all the boxes of the first row of the
tableaux $Y_\alpha$. ${\cal V}^{-1}=(1-T_1)(1-T_2)$ is a sort of
"volume factor" 
\footnote{For small $\epsilon_1, \epsilon_2$, ${\cal V}^{-1}\sim\epsilon_1\epsilon_2$.
We can be cavalier about the renormalization of the prefactor
since none of the conclusions of this paper depend on it.
A careful study of the renormalization is nevertheless desirable.}
that we will  always omit in the following. 
We stress that the factor ${\cal V}^{-1}$ is needed in order to
ensure that no negative terms appears in the expansion of the
(\ref{lmnfinal}) so to preserve the interpretation as a Chern character. 
The numerator in
(\ref{lmnfinal}) reproduces the result in \cite{Losev:2003py}.
The 0-form part $\tr\tilde\varphi_{\rm bos}^m $ evaluated at the
critical point corresponding to the Young diagrams $\{Y_\alpha\}$,
is now given by expanding the exponentials in (\ref{lmnfinal}) and
taking their $m$-th power \beqa {\cal O}_m(\{
Y_\alpha\})&\equiv&\tr \tilde\varphi_{\rm
bos}^m\vert_{\{Y_\alpha\}}
=\sum_{\alpha=1}^n\bigg\{
a_\alpha^m+\sum_{j_\alpha=1}^{k_{1_\alpha}}\bigg[[a_\alpha+
\epsilon_2(j-1)+\epsilon_1k^\prime_{j_\alpha}+]^m
\nonumber\\
&& -[a_\alpha-\epsilon_2(j_\alpha-1)]^m-[a_\alpha+\epsilon_2
j_\alpha+\epsilon_1k^\prime_{j_\alpha})]^m+ [a_\alpha+\epsilon_2
j_\alpha]^m\bigg]\bigg\} \label{fintrfim}
\eeqa
Finally
\beq
\langle\tr \tilde\varphi^m\rangle={1\over {\cal Z} } \sum_{\{k;Y_\alpha
 \}} {\sum_{\gamma=1}^n {\cal
O}_m(\{ Y_\alpha\})\over \prod_{\alpha,\beta}^n \prod_{s\in
Y_{\alpha}} E_{\alpha\beta}(s)(2\epsilon_+-E_{\alpha\beta}(s))} q^k
\label{recipe}
\eeq
with
the sum running over all the sets of $n$ tableaux $Y_\alpha$ with $
\sum_\alpha\vert Y_\alpha\vert=k$.

\subsection{Matone's relation}
In this section we verify that it is $\langle\tr
\tilde{\varphi}^2\rangle$ rather than $\langle\tr
\varphi^2\rangle$, which satisfies the Matone's relation. As we
have already mentioned  the two expressions are the same in the
limit in which $\epsilon_{1,2}\to 0$, they only differ when
gravitational corrections are turned on.

Introducing a coupling $\tau_m$ we define \beq
e^{-{1\over\epsilon_1\epsilon_2 } {\cal F}(\tau,\tau_m) }=\int
{\rm exp} (-S-\tau_m \tr\tilde{\varphi}^m). \label{poteff} \eeq
Then ${\cal Z}=\exp (-{\cal F}(\tau,0)/h^2)$. As we already said
$\tr \tilde\varphi_{(0)}^m\vert_{\{Y_\alpha\}}$ can be found by
expanding the exponential in the numerator of (\ref{lmnfinal}).
The first non-trivial term appears at $m=2$ \beq \tr
\tilde\varphi_{(0)}^2\vert_{\{Y_\alpha\}}=a^2+k \eeq Only in this
case the result does not depends on the shape of the diagram but
only on the total winding number $k$. One finds
\beqa
<\tr\tilde{\varphi}^2>&\equiv&\sum_k{\cal G}_kq^k\equiv {1\over
{\cal Z}}{\partial{\cal F}(\tau,\tau_2)\over\partial\tau_2}
\vert_{\tau_2=0}\nn\\
&=&{1\over {\cal Z}}\sum_k (a^2+\epsilon_1\epsilon_2k)\,q^k\,{\cal Z}_k =
a^2+\epsilon_1\epsilon_2 q{\partial\ln {\cal Z}\over\partial q}=a^2+q{\partial
{\cal F}(\tau,0)\over\partial q}\nonumber\\
&=&a^2+\sum_k k\,{\cal F}_k\, q^k.
\label{formfinal}
\eeqa
${\cal G}_k=k{\cal F}_k$ is Matone's relation. This result obviously
extends to $N=2^*$ and to the case of matter in the fundamental.
The only difference in these cases is the value of the
determinants in ${\cal Z}_k$. See \cite{Bruzzo:2002xf} for their
explicit form in these cases. Finally, (\ref{formfinal}) holds at
all orders in powers of $\epsilon_1, \epsilon_2$ meaning that the
Matone's relation should hold once gravitational corrections are
taken into account.

\subsection{The chiral ring}
Here we show that the chiral ring relations
are satisfied by $\tilde\varphi$.
We refer to \cite{Cachazo:2002ry} for notations and detailed explanations.
The Coulomb branch of the $U(n)$
gauge theory with $N=2$ supersymmetry can be parametrized by the
$n$ gauge invariant expectation values
\begin{equation}
u_k=\langle Tr \Phi^k\rangle.
\end{equation}
It is known that the expectation values $\langle Tr
\Phi^m\rangle$ could be expressed via
$u_1,\cdots,u_n$. This relationship could most conveniently  be
written as
\begin{equation}
\langle Tr \frac{1}{z-\Phi}\rangle
=\frac{P_n^\prime(z)}{\sqrt{P_n(z)^2-4 q}}\, \, ,
\label{ringrelation}
\end{equation}
where $P_N(z)$ is the polynomial defining
the quantum Seiberg-Witten curve. (\ref{ringrelation}) gives an
infinite number of relations, which can be obtained by expanding
both sides into powers of $1/z$. In particular, when the gauge
group is $SU(2)$, we have only one independent expectation value
$u\equiv u_2=Tr \Phi^2$. In this case (\ref{ringrelation})
gives
\begin{eqnarray}
\langle Tr \Phi^4\rangle &=&4 q+\frac{u^2}{2} \, , \nonumber \\
\langle Tr \Phi^6\rangle &=&6qu + \frac{u^3}{4} \, , \nonumber \\
\langle Tr \Phi^8\rangle &=&12 q^2 + 6 q u^2 + \frac{u^4}{8}
\label{su2ringrelations}
\end{eqnarray}
Using(\ref{recipe}) we have computed $\langle\tr
\tilde\varphi^m\rangle$ for $m\le 8$ up to 4 instanton
contribution\footnote{ We did this with Mathematica. We have
checked higher instantons numbers and powers of the scalar field
too. Here we write few sample expressions that can fit the page.}.
The result reads:
\begin{eqnarray}
\langle \tr \tilde\varphi^8\rangle =
2a^8\left[1+\frac{14q}{a^4}+\frac{161q^2}{8a^8}+
\frac{35q^3}{4a^{12}}+\frac{15337q^4}{2048a^{16}}\right.
\nonumber \\
+\left(\frac{105q}{2a^6}+\frac{497q^2}{16a^{10}}+
\frac{1505q^3}{64a^{14}}+ \frac{99561q^4}{2048a^{18}}\right)
(\epsilon_1^2+\epsilon_2^2) \nonumber \\
\left. +\left(\frac{70 q}{a^6}+\frac{35q^2}{a^{10}}+
\frac{469q^3}{16a^{14}}+ \frac{17073q^4}{256a^{18}}\right)
\epsilon_1 \epsilon_2 +\cdots \right] ,  \nonumber \\
\langle \tr\tilde\varphi^6\rangle =2a^6
\left[1+\frac{15q}{2a^4}+\frac{135q^2}{32a^8}+
\frac{125q^3}{64a^{12}}+ \frac{16335q^4}{8192a^{16}}  \right. \nonumber \\
+\left(\frac{75q}{8a^6}+\frac{135q^2}{64a^{10}}+\frac{735q^3}{128a^{14}}+
\frac{124575q^4}{8192a^{18}}\right)(\epsilon_1^2+\epsilon_2^2)
\nonumber \\
\left.
+\left(\frac{45q}{4a^6}+\frac{75q^2}{32a^{10}}+\frac{525q^3}{64a^{14}}+
\frac{92175q^4}{4096a^{18}}\right)\epsilon_1 \epsilon_2 +\cdots \right], \nonumber \\
\langle \tr\tilde\varphi^4\rangle= 2 a^4 \left[
1+\frac{3q}{a^4}+\frac{9q^2}{16a^8}+
\frac{7q^3}{16a^{12}}+\frac{2145q^4}{4096a^{16}}  \right.
\nonumber \\
+\left(\frac{q}{4a^6}+\frac{25q^2}{32a^{10}}+
\frac{267q^3}{128a^{14}}+
\frac{22529q^4}{4096a^{18}}\right)(\epsilon_1^2+ \epsilon_2^2)
\nonumber \\
\left. +\left(\frac{9q^2}{8a^{10}}-\frac{3q^3}{16a^{14}}+
\frac{8535q^4}{1024a^{18}}\right) \epsilon_1 \epsilon_2 +\cdots \right], \nonumber  \\
\langle \tr\tilde\varphi^2\rangle= 2a^2
\left[1+\frac{q}{2a^4}+\frac{5q^2}{32a^8}
+\frac{9q^3}{64a^{12}}+ \frac{1469q^4}{8192a^{16}}\right. \nonumber \\
+\left(\frac{q}{8a^6}+\frac{21q^2}{64a^{10}}+
\frac{55q^3}{64a^{14}}+\frac{18445q^4}{8192a^{18}}\right)(\epsilon_1^2+
\epsilon_2^2) \nonumber \\
\left. +\left(\frac{q}{4a^6}+\frac{19q^2}{32a^{10}}+
\frac{47q^3}{32a^{14}}+\frac{15151q^4}{4096a^{18}}
\right)\epsilon_1 \epsilon_2 +\cdots \right].
\label{PhiJexpansion}
\end{eqnarray}
In the limit $\epsilon_{1,2}\to 0$, (\ref{PhiJexpansion})
satisfy the ring relations (\ref{su2ringrelations}) up to
order $q^4$.
The $\epsilon$-terms represent the first gravitational
corrections to the chiral ring. It would be nice to reproduce
these deformations of the chiral ring relations from the matrix model
perspective (may be along the lines of \cite{David:2003ke})
\footnote{$\epsilon$-terms in $\tr \tilde{\varphi}^2$ have been shown to match
the instanton corrections to gravitational couplings in the ${\cal N}=2$
gauge theory \cite{Klemm:2002pa}.}.

\section*{Acknowledgements}
F.F. wants to thank A. Losev for patiently explaining to him the results he
obtained with his collaborators.
J.F.M thanks J.R. David, E. Gava and K.S. Narain for useful discussions.
R.F. and R.P. have been partially supported by the Volkswagen foundation
of Germany. 
R.P. wants to thank I.N.F.N. for supporting
a visit to the University of Tor Vergata.
This work was supported in part by the EC contract
HPRN-CT-2000-00122, the EC contract HPRN-CT-2000-00148, 
the EC contract HPRN-CT-2002-00325,
the MIUR-COFIN contract 2003-023852, the NATO
contract PST.CLG.978785 and the INTAS contracts 03-51-6346 and 00-561.
\appendix

\section{Appendix}
\setcounter{equation}{0}
In this appendix we discuss
the "commutative" limit $\zeta \rightarrow 0$ of matrix elements essential
for connecting the correlators computed in the non commutative case with
those computed in the commutative case. We hope that following this line
of reasoning it will be possible to calculate the renormalizing "volume" factor
mentioned at the end of section (4.1). This should  be needed to
establish chiral ring relations in the $\epsilon$ deformed case.

Let us suppose that the solution to the equation
\begin{equation}
\bar{\Delta}(\bar z_0,z_0)U_{cl}(z_0,\bar z_0)=0
\end{equation}
in the commutative case ($\zeta=0$) is known\footnote{To simplify the notation,
in this appendix $z=(z_1,z_2)$. In agreement with the main text, $\zeta$ plays
the role of $\hbar$.}.
The function
\begin{equation}
U(z_0,\bar z_0|z)=U_{cl}(z_0,\bar z_0)e^{
\frac{\bar z_0z}{\zeta}-\frac{\bar{z_0} z}{2\zeta}}
\end{equation}
in the quasi-classical limit $\zeta\rightarrow 0$
satisfies the "noncommutative equation"
\begin{equation}
\bar{\Delta}(z,\bar{z})U(z_0,\bar z_0|z)=0,
\end{equation}
in leading order
where $z$ and $\bar{z}$ are the non-commutative coordinates,
$[\bar{z},z]=\zeta$ (in the holomorphic representation, the states are
holomorphic functions of $z$ and $\bar{z}=\zeta\partial/\partial
z$ acts as a derivative). Up to a normalization, the wave function
\begin{equation}
e^{ \frac{\bar z_0 z}{\zeta}-\frac{\bar z_0 z_0}{2\zeta}}
\label{coherentstate}
\end{equation}
is the wave packet, centered around the classical variables $(z_0,\bar z_0)$,
which minimizes Heisenberg's uncertainty relations i.e. a coherent state.
Let us calculate the generic matrix
element\footnote{Zero modes written in terms of the ADHM variables are always
sandwiched between a pair of $U(z)$'s.}
\begin{equation}
\langle z_0 |\varphi |\bar z_0^\prime\rangle \equiv\int
\bar{U}(\bar{z_0},z_0|\bar{z}){\cal A} U(z_0^\prime,\bar
z_0^\prime|z)e^{-\frac{z\bar{z}}{\zeta}}
\frac{d^4z}{(\pi\zeta)^2}=\bar{U}_{cl}(\bar z_0,z_0){\cal
A}U_{cl}(z_0^\prime,\bar{z_0}^\prime) e^{\frac{\bar z_0^\prime
z_0}{\zeta}-\frac{\bar z_0 z_0+\bar z_0^\prime
z_0^\prime}{2\zeta}}
\end{equation}
Now as the simplest example we calculate $\tr \varphi^2$: \beqa
&&\tr \int \langle z_0 |\varphi |\bar{z_0^\prime}\rangle \langle
z_0^\prime |\varphi |\bar{z_0}\rangle \frac{d^4z_0^\prime d^4
z_0}{(\pi \zeta)^4}
\nonumber \\
&=&\tr \int \bar{U}_{cl}(\bar{z_0},z_0){\cal
A}U_{cl}(z_0^\prime,\bar{z_0}^\prime)
\bar{U}_{cl}(\bar{z_0}^\prime,z_0^\prime){\cal A}U_{cl}(z_0,\bar{z_0})
e^{-\frac{|z_0-z_0^\prime|^2}{\zeta}} \frac{d^4z_0^\prime d^4z_0}{(\pi
\zeta)^4} \nonumber \\
&\approx& \tr \int \bar{U}_{cl}(\bar{z_0},z_0){\cal
A}U_{cl}(z_0,\bar{z_0}) \bar{U}_{cl}(\bar{z_0},z_0){\cal
A}U_{cl}(z_0,\bar{z_0}) \frac{d^4 z_0}{(\pi \zeta)^2}. 
\eeqa For small
$\zeta$, $\exp-\frac{|z_0-z_0^\prime|^2}{\zeta}/(\pi \zeta)^2$ can
be approximated by $\delta^{4}(z_0-z_0^\prime)$. Note that in the
first line of the above equation the usual insertions of
exponential measure factors are absent since they are already
distributed in the wave functions (see (\ref{coherentstate})). So
for $\zeta\to 0$ we can approximate expectation values with
integrals over the classical variables. A similar result holds
also for the higher powers of $\varphi$.

\end{document}